\documentclass[final,twocolumn]{elsarticle}
\usepackage{amssymb}
\journal{Physics Letters B}
\begin{document}
\begin{frontmatter}

\title{Neutron/proton ratio of nucleon emissions as a probe of neutron skin}

\author[a,b]{X. Y. Sun}
\author[a]{D. Q. Fang}\ead{dqfang@sinap.ac.cn}
\author[a]{Y. G. Ma}\ead{ygma@sinap.ac.cn}
\author[a]{X. Z. Cai}
\author[a]{J. G. Chen}
\author[a]{W. Guo}
\author[a]{W. D. Tian}
\author[a]{H. W. Wang}
\address[a]{Shanghai Institute of Applied Physics,
Chinese Academy of Sciences, Shanghai 201800, China}
\address[b]{Graduate School of Chinese Academy of Sciences,
Beijing 100039, China}

\begin{abstract}
The dependence between neutron-to-proton yield ratio ($R_{np}$) and
neutron skin thickness ($\delta_{np}$) in neutron-rich projectile
induced reactions is investigated within the framework of the
Isospin-Dependent Quantum Molecular Dynamics (IQMD) model. The
density distribution of the Droplet model is embedded in the
initialization of the neutron and proton densities in the present
IQMD model. By adjusting the diffuseness parameter of neutron
density in the Droplet model for the projectile, the relationship
between the neutron skin thickness and the corresponding $R_{np}$ in the
collisions is obtained. The results show strong linear correlation
between $R_{np}$ and $\delta_{np}$ for neutron-rich Ca and Ni
isotopes. It is suggested that $R_{np}$ may be used as an
experimental observable to extract $\delta_{np}$ for neutron-rich
nuclei, which is very significant to the study of the nuclear structure
of exotic nuclei and the equation of state (EOS) of
asymmetric nuclear matter.
\end{abstract}

\end{frontmatter}


   Nuclear radius is one of the basic quantities of a nucleus.
The proton root-mean-square (RMS) radius can be determined to very
high accuracy via the charge radius measured by electromagnetic
interactions, typically with an error of 0.02 fm or better for
many nuclei~\cite{Fricke}. In contrast, it is much more difficult
to accurately determine the neutron density distribution of a
nucleus experimentally~\cite{Ray}. Thus, the accuracy
of experimental neutron radius is much lower than that of the
proton radius. However, the information of neutron density is very
important to the study of nuclear structure for neutron-rich
nuclei, atomic parity non-conservation, iso-vector interactions,
and neutron-rich matter in astrophysics etc. It is remarkable that
a single measurement has so many applications in the research
fields of atomic, nuclear and astrophysics~\cite{Horowitz,
Danielewicz}.

   A nucleus with neutron number ($N$) being larger than proton
number ($Z$) are expected to have a neutron skin (defined as the
difference between the neutron and proton RMS radii:
$\delta_{np}\equiv \langle r^{2}_{n} \rangle ^{1/2}- \langle
r^{2}_{p} \rangle^{1/2} )$. The neutron skin thickness
$\delta_{np}$ depends on the balance between various aspects of
the nuclear force. The actual proton and neutron density
distributions are determined by the balance between the isospin
asymmetry and Coulomb force. $\delta_{np}$ is found to be related
with a constraint on the equation of state (EOS) of asymmetric nuclear
matter. Strong linear correlation between $\delta_{np}$ and
$L$ (the slope of symmetry energy coefficient $C_{sym}$),
the ratio $L/J$ ($J$ is the symmetry energy coefficient at
the saturation density $\rho_0$), $J-a_{sym}$
($a_{sym}$ is the symmetry energy coefficient of finite nuclei)
are demonstrated~\cite{nskin-prl}.
This constraint is important for extrapolation of
the EOS to high density and hence useful for studying properties
of neutron star~\cite{nskin-prl,nskin-Esym,Chen L W,
Brown,Yoshida,Suzuki T, Sam,Liu}. Neutron skin thickness can yield
a lot of information on the derivation of volume and surface
symmetry energy, as well as nuclear incompressibility with
respective to density. $\delta_{np}$ is significant to the study
of the EOS in different theoretical models such as Skyrme Hartree-Fock
(SHF)~\cite{Horowitz,Danielewicz,Chen L W,Brown,Yoshida,Sam,Liu},
relativistic mean-field (RMF)~\cite{Yoshida,Suzuki T,Bhagwat}, BUU
model~\cite{Ma1,Ma2,Cai X Z} and Droplet model~\cite{myers}.
Furthermore, neutron skin thickness helps to identify a nucleus
with exotic structure. Thus the precise determination of $\delta_{np}$
for a nucleus becomes an important research subject in nuclear physics.

Several attempts have been made or suggested to determine the
neutron density distribution such as using proton scattering,
interaction cross sections in heavy ion collisions at relativistic
energies \cite{Suzuki}, parity violating measurements
\cite{Horowitz,Danielewicz,Chen L W,Brown,Yoshida}, neutron
abrasion cross sections in heavy ion collisions~\cite{Chunwang
Ma}. On the other hand, the neutron and proton transverse emission
and double neutron-proton ratios have been studied as a sensitive
observable of the asymmetry term of the nuclear EOS in the
experiment \cite{MSU} and different kinds of simulations
\cite{Ma-lgm,ZhangYX-PLB,Tsang09,Li,Toro}. In this Letter, the
relationship between $\delta_{np}$ and the ratio of the emitted
neutron and proton yields ($R_{np}=Y_n/Y_p$) in neutron-rich
projectile induced reactions is, for the first time, presented
within the framework of Isospin-Dependent Quantum Molecular
Dynamics (IQMD) model. The possibility of extracting $\delta_{np}$
from $R_{np}$ is investigated.

The QMD approach is a many-body theory that describes heavy-ion
reactions from intermediate to relativistic energies~\cite{J.Aichelin}.
It includes several important factors:
initialization of the projectile and target, nucleon propagation
in the effective potential, nucleon-nucleon (NN) collisions in a nuclear
medium and the Pauli blocking effect. A general review of
the QMD model can be found in \cite{Aichelin}. The IQMD model is
based on the QMD model with consideration of the isospin effect.

  The dynamics of heavy ion collision at intermediate
energies is governed mainly by three components: the mean field,
two-body collisions, and Pauli blocking. Therefore, for an
isospin-dependent reaction dynamics model, it is important to include
isospin degrees of freedom with the above three components. In
addition, the sampling of phase space of neutrons and protons in the
initialization should be treated separately because of the large
difference between neutron and proton density distributions for
nuclei that far from the $\beta$-stability line.

In the IQMD model, each nucleon
$i$ is represented by a Gaussian wave packet with definite width
($L=2.16$ fm$^{2}$) centered around the mean position and the mean
momentum:
\begin{eqnarray}
\Psi_{i}(\overrightarrow{\bold{r}},t)=\frac{1}{(2\pi L)^{3/4}}
 & \exp[-\frac{(\overrightarrow{\bold{r}}-\overrightarrow{\bold{r}_{i}}{(t)})^{2}}{4L}] \nonumber  \\
 & \times\exp[\frac{i\overrightarrow{\bold{r}}\cdot\overrightarrow{\bold{p}_{i}}{(t)}}{\hbar}]
\end{eqnarray}
The nuclear mean field in the IQMD model is determined as following
\begin{eqnarray}
U(\rho,\tau_{z}) =&\alpha(\frac{\rho}{\rho_{0}})+\beta(\frac{\rho}{\rho_{0}})^{\gamma}
                   +\frac{1}{2}(1-\tau_{z})V_{c} \nonumber \\
                  & +C_{\mathrm{sym}}\frac{(\rho_{n}-\rho_{p})}{\rho_{0}}\tau_{z}+U^\mathrm{Yuk}
\end{eqnarray}
with the normal nuclear matter density $\rho_{0}=0.16$~fm$^{-3}$.
$\rho$, $\rho_{n}$ and $\rho_{p}$ are the total, neutron and
proton densities, respectively. $\tau_{z}$ is $z$-th component of
the isospin degree of freedom, which equals $1$ or $-1$ for neutron
or proton, respectively. The coefficients $\alpha$, $\beta$ and
$\gamma$ are parameters for the nuclear EOS.
$C_\mathrm{sym}$is the symmetry energy strength due to the
difference between neutron and proton. In the present work, we take
$\alpha=-356$ MeV, $\beta$ = 303 MeV and $\gamma = 1.17$ which
correspond to the so-called soft EOS with $C_\mathrm{sym} = 32$~MeV
and incompressibility of $K=200$~MeV. $V_\mathrm{c}$ is the
Coulomb potential and $U^\mathrm{Yuk}$ is Yukawa (surface)
potential. In the phase space initialization of the projectile and
target, the density distributions of proton and neutron are
distinguished from each other. The neutron and proton density
distributions for the initial projectile and target nuclei in the
present IQMD model are taken from the Droplet Model \cite{Myers}.

\begin{figure}[t]
\centering\includegraphics[width=6.5cm]{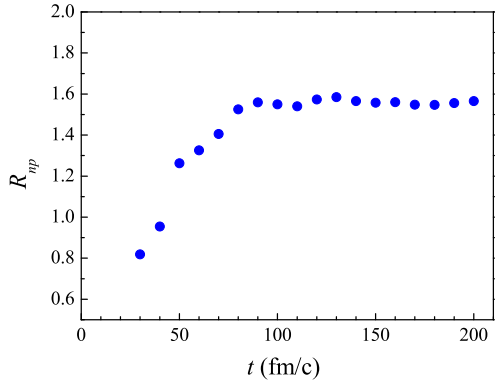}
\caption{\label{fig1}(Color online.) The evolution time dependence of $R_{np}$
for $^{50}$Ca+$^{12}$C at 50A MeV.}
\end{figure}

In the Droplet model, we can change the diffuseness parameter to
get different density,
\begin{eqnarray}
\rho_{i}(r)=\frac{\rho^{0}_{i}}{1+\exp(\frac{r-C_{i}}{f_{i}t_{i}/4\cdot4})},i=n,p
\end{eqnarray}
where $\rho^{0}_{i}$ is the normalization constant which ensures that the
integration of the density distribution equals to the number of neutrons (i=n)
or protons (i=p); $t_{i}$ is the diffuseness parameter; $C_{i}$ is the
half density radius of neutron or proton determined by the
Droplet model \cite{Myers}
\begin{eqnarray}
C_{i}=R_{i}[1-(b_{i}/R_{i})^{2}],i=n,p
\end{eqnarray}
here $b_{i}=0.413f_{i}t_{i}$, $R_{i}$ is the equivalent sharp
surface radius of neutron and proton. $R_{i}$ and $t_{i}$ are
given by the Droplet model. A factor $f_{i}$ is introduced by us
to adjust the diffuseness parameter. In \cite{Trzci'nska},
Trzci'nska et al. found that the half density radii for neutrons
and protons in heavy nuclei are almost the same, but the diffuseness
parameter for neutron is larger than that for the proton. So we
introduce the factor  $f_{i}$ to adjust the diffuseness parameter
for neutron. In the calculation for neutron-rich nucleus,
$f_{p}=1.0$ is used in Eq.(3) for the proton density distribution
as in the Droplet model, while $f_{n}$ in Eq.(3) is changed from
1.0 to 1.6. Different values of $\delta_{np}$ will be deduced from
Eq.(3). Using the density distributions of the Droplet model, we
can get the initial coordinate of nucleons in nuclei in terms of
the Monte Carlo sampling method. The momentum distribution of
nucleons is generated by means of the local Fermi gas
approximation:
 \begin{eqnarray}
P^{i}_{F}(\overrightarrow{\bold{r}})=\hbar[3\pi^{3}\rho_{i}(\overrightarrow{\bold{r}})]^{1/3}
 ,(i=n,p)
 \end{eqnarray}
 In the IQMD model, the nucleon's radial density can be written as
\begin{eqnarray}
 \rho(r)=&\sum\frac{1}{(2\pi L)^{3/2}}\exp(-\frac{r^{2}+r_{i}^{2}}{2L})
              \frac{L}{2rr_{i}}[\exp(\frac{rr_{i}}{L})\nonumber
              \\
         &-\exp(-\frac{rr_{i}}{L})]
\end{eqnarray}

To avoid taking an unstable initialization of projectile and target
in the IQMD calculation, we only select the initialization samples of
those nuclei that meet the required stability conditions.
The binding energies and root mean square (rms) radii of
the initial configuration for the colliding nuclei are required
to be stable. Using the selected initialization phase
space of nuclei in IQMD to simulate the collisions,
the nuclear fragments are constructed by a modified isospin-dependent
coalescence model, in which nucleons with relative momentum smaller
than $P_{0}=300$~MeV/c and relative distance smaller than
$R_{0}=3.5$~fm will be combined into a cluster.

\begin{figure}[tb]
\centering\includegraphics[width=6.5cm]{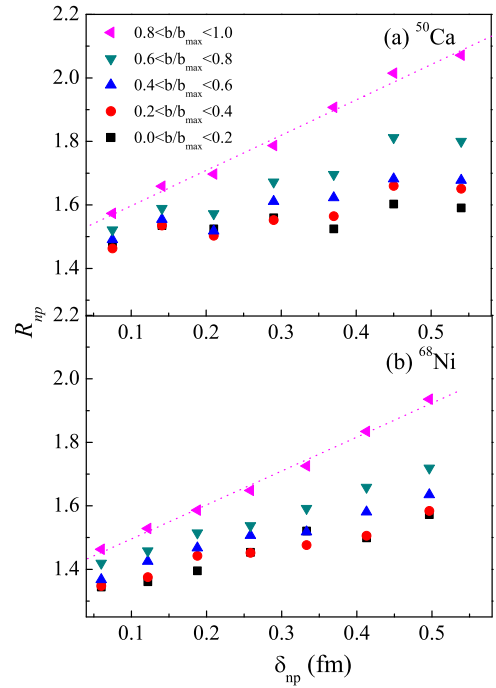}
\caption{\label{fig2}(Color online.)
The dependence of $R_{np}$ on neutron skin thickness
of the projectile for $^{50}$Ca+$^{12}$C (a) and $^{68}$Ni+$^{12}$C (b)
at 50A MeV under the condition of $Y>0$. Different symbols are used for
different range of the reduced impact parameter as shown in the legend.
The dotted lines just guide the eye.}
\end{figure}

\begin{figure}[tb]
\centering\includegraphics[width=6.5cm]{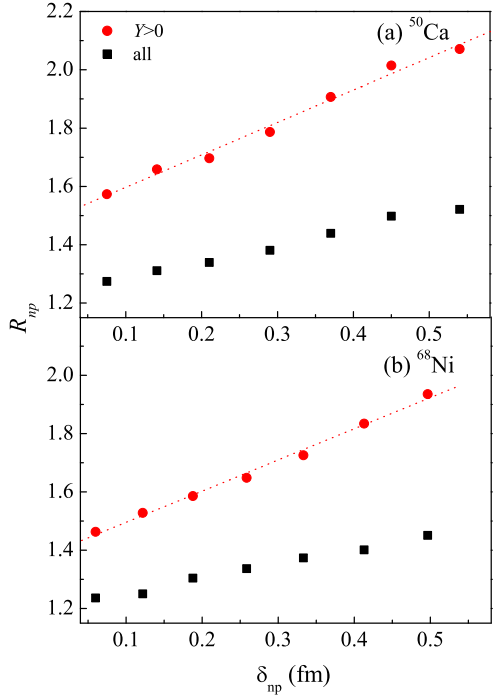}
\caption{\label{fig3}(Color online.)
The dependence of $R_{np}$ on neutron skin thickness
under the condition of reduced impact parameter from 0.8 to 1.0
for $^{50}$Ca+$^{12}$C (a) and $^{68}$Ni+$^{12}$C (b) at 50A MeV.
The results without any gate on the rapidity ($Y$) are shown as solid squares.
The results with $Y>0$ are plotted as solid circles.
The dotted lines just guide the eye.}
\end{figure}


The collision processes of some Ca and Ni isotopes with $^{12}$C
target at 50A MeV are simulated using the IQMD model. The
fragments including neutrons and protons that formed during the evolution
of the collision are constructed by the coalescence method. The
yield ratio $R_{np}$  of the emitted neutrons and protons can be
calculated from the yields of the produced neutrons and protons.
By changing the factor $f_{n}$ in the neutron density distribution
of the Droplet model for the projectile, different values of
$\delta_{np}$ and the corresponding $R_{np}$ are obtained. Thus we
can obtain the correlation between $R_{np}$ and $\delta_{np}$. In the
calculation, the time evolution of the dynamical process was
simulated until $t=200$~fm/c. The calculated $R_{np}$ is stable after
150 fm/c as shown in Fig.~\ref{fig1}, so we accumulate the
emitted neutrons and protons between 150 fm/c and 200 fm/c in order
to improve statistics. The $R_{np}$ from different range of the
reduced impact parameter for $^{50}$Ca +$^{12}$C and $^{68}$Ni
+$^{12}$C are plotted in Fig.~\ref{fig2}. The reduced impact
parameter is defined as $b/b_{\mathrm{max}}$ with
$b_{\mathrm{max}}$ being the maximum impact parameter. From
Fig.~\ref{fig2}, we can see that $R_{np}$ rises as $\delta_{np}$
increases. With $\delta_{np}$ being fixed, $R_{np}$ also
becomes larger with the increasing of the reduced impact parameter. It
means that $R_{np}$ from peripheral collisions is usually larger than
that from central collisions. The main purpose of the present
study is to investigate the relationship between $R_{np}$ and the
neutron skin thickness of the projectile. In order to minimize the
target effect on $R_{np}$, we use rapidity
($Y$) cut to select neutrons and protons from the projectile. The
rapidity of the fragment which is normalized to the incident
projectile rapidity is defined as:
\begin{eqnarray}
Y=\frac{1}{2}\mathrm{log}(\frac{E + p_{z}}{E -
p_{z}})/Y_\mathrm{proj} ,
\end{eqnarray}
where $E$ is the energy of the fragment, $p_{z}$ is the momentum
of $z$ direction, $Y_\mathrm{proj}$ is the rapidity of the
projectile. We choose $Y>0$ to strip away fragments coming from
the target. From the results shown in Fig.~\ref{fig3}, we can see
that $R_{np}$ with $Y>0$ are larger than $R_{np}$ without $Y$ cut.
Since the projectile is neutron-rich nucleus, the correlation between
$R_{np}$ and $\delta_{np}$ is stronger for nucleons coming from the
projectile than those from both the projectile and target.

\begin{figure}[tb]
\centering\includegraphics[width=6.5cm]{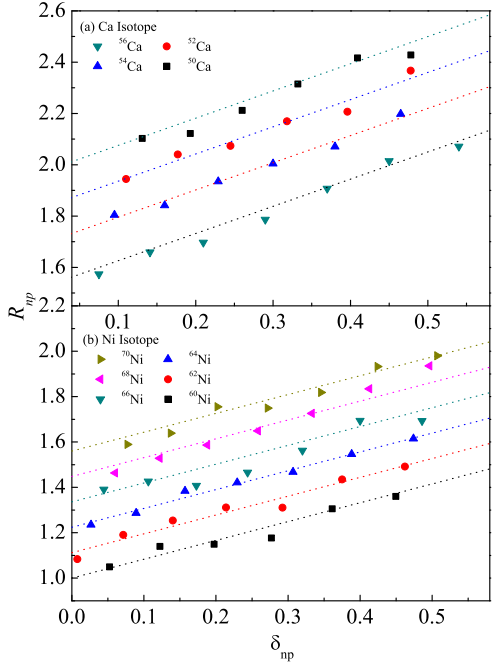}
\caption{\label{fig4}(Color online.)
The dependence of $R_{np}$ on neutron skin thickness
under the condition of Y $>$0 and $0.8<b/b_{\mathrm{max}}<1.0$
for $^{50,52,54,56}$Ca+$^{12}$C (a) and
$^{60,62,64,66,70}$Ni+$^{12}$C (b) at 50A MeV. Different symbols
are used for different projectile as shown in the legend.
The dotted lines just guide the eye, for detail see text.}
\end{figure}

\begin{figure}[tb]
\centering\includegraphics[width=6.5cm]{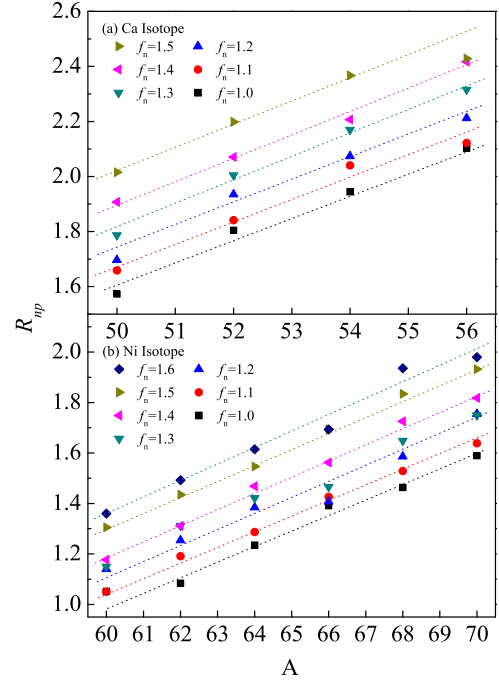}
\caption{\label{fig5}(Color online.)
The dependence of $R_{np}$ on the mass number $A$
under the condition of $Y>0$ and $0.8<b/b_{\mathrm{max}}<1.0$
for $^{50,52,54,56}$Ca+$^{12}$C and $^{60,62,64,66,70}$Ni+$^{12}$C
at 50A MeV. Different symbols are used for different value of the
factor $f_n$ as shown in the legend. The dotted lines just guide
the eye, for detail see text.}
\end{figure}

Strong linear correlation between $R_{np}$ and $\delta_{np}$ of the
projectile, especially for $0.8<b/b_{\mathrm{max}}<1.0$, is
exhibited in Fig.~\ref{fig2}. It indicates that $R_{np}$ is very
sensitive to $\delta_{np}$ of the projectile, especially in
peripheral collisions. For systematic study, reactions of other Ca
and Ni isotopes such as $^{52,54,56}$Ca and $^{60,62,64,66,70}$Ni
are also calculated. The results with the reduced impact parameter
from 0.8 to 1.0 and the rapidity being positive are shown in
Fig.~\ref{fig4}. Strong linear relationship between $R_{np}$ and
$\delta_{np}$ are also observed for all projectiles. From
Fig.~\ref{fig4}, a linear function ($R_{np}=a+b\cdot\delta_{np}$) can
describe the correlation between $R_{np}$ and $\delta_{np}$ well for
both Ca and Ni isotopes. The dependence between $R_{np}$ and
$\delta_{np}$ is fitted using this linear function. The mean
slopes for Ca and Ni isotopes are 1.06 and 0.83, respectively.

From the above discussions, $R_{np}$ could be viewed  as a sensitive
observable of $\delta_{np}$ for the projectile. If the produced
neutrons and protons are measured by experiment, it is possible to
extract $\delta_{np}$ from the neutron-to-proton yield ratio
$R_{np}$. From the fitted slope values, the estimated error of
$\delta_{np}$ may be less than 0.08 fm if the uncertainty of
$R_{np}$ is less than $5\%$.
It should be pointed out that this uncertainty does not take
into account the error arising from the determination of impact parameter,
which is always an important issue in studies of nuclear reactions
in the energy range of our calculation.
We estimate the uncertainty of $\delta_{np}$ for
$0.6<b/b_{\mathrm{max}}<1.0$ to be around 0.1 fm when the uncertainty
of $R_{np}$ is $5\%$.
To achieve high resolution, well defined experimental determination
of the centrality of the reaction is required for this method.

In our calculation, different $\delta_{np}$ is obtained by
changing the factor $f_n$ in Eq.(3). We can also study the mass
dependence of $R_{np}$ with the same $f_n$ as shown in
Fig.~\ref{fig5}. For $f_n=1.0$, it refers to the neutron skin
thickness of the Droplet model's prediction. In this figure, the linear
dependence of $R_{np}$ with the mass number $A$ is also seen at
different value of $f_n$. Due to the acceptance and efficiency in
measurement of the emitted neutron and proton, the uncertainty of $R_{np}$
may be not easy to be estimated which makes it difficult to obtain
the error of $\delta_{np}$. There is one method to minimize
these effects:  we can measure $R_{np}$ of one isotope chain with
different mass number. In this case, the acceptance and efficiency of neutron
and proton will not change the mass dependence of $R_{np}$. If
$\delta_{np}$ of some nuclei is larger than the normal value of
the Droplet model, it will deviate from the line of $f_n=1.0$.
By studying the mass dependence of $R_{np}$, it is possible to
extract $\delta_{np}$ and identify nucleus with abnormal
neutron skin thickness.

The neutron excess ($I=(N-Z)/A$) dependence of neutron skin thickness
for four parameter sets (NL3, NL-SH, SkM*, SIII) in the Droplet model
and self-consistent extended Thomas-Fermi (ETF) calculations have been
investigated by Warda {\it et al}.~\cite{Warda}. Differences of the
neutron skin thickness among different predictions increase with the
increase of $I$. To distinguish which potential parameter set or method
is more close to the experimental data, higher resolution measurement on
neutron skin thickness is required for stable nuclei having small $I$
in comparison with nuclei far from the stability line having large $I$
value. From their results, the neutron skin thickness by different
calculation varies from 0.01 fm to 0.11 fm at $I=0.11$.
In this sense, it will be capable of distinguishing different theoretical
predictions when $I>0.11$ by extracting $\delta_{np}$ from $R_{np}$
measurement which the present paper proposes. In other words, the present
method will be more feasible for heavy nuclei or intermediate mass nuclei
far from the $\beta$-stability line.

  In summary, we have made calculation on the relationship between $R_{np}$
and the neutron skin thickness for the first time.
The simulated data for Ca and Ni isotopes induced reactions
show a good linear correlation
between $R_{np}$ and $\delta_{np}$ for neutron-rich
projectile. It is suggested that $R_{np}$ could be used as an
experimental observable to extract the neutron skin thickness
for neutron-rich nucleus. When the neutron skin thickness of
one isotope is obtained, we can get some information on the
EOS in different theoretical models such as Skyrme Hartree-Fock,
relativistic mean-field and BUU model. From the isotope
dependence of neutron skin thickness, a lot of information on the
derivation of volume and surface symmetry energy, as well as nuclear
incompressibility with respective to density can be extracted.
This is very important to the study of the EOS of asymmetric nuclear matter.

\section*{Acknowledgments} \nonumber
This work is supported by National Natural Science Foundation
of China under contract No.s 10775168, 10775167, 10979074 and 10747163,
Major State Basic Research Development Program in China under
contract No. 2007CB815004, Knowledge Innovation Project of
Chinese Academy of Sciences under contract No. KJCX3.SYW.N2,
and the Shanghai Development Foundation for Science and Technology
under contract No. 09JC1416800.

\end{document}